\documentclass{ws-procs9x6}

\newcommand{\un}[1]{\,\mathrm{#1}}

\begin{document}

\title{STATUS OF THE ICETOP AIR SHOWER ARRAY AT THE SOUTH POLE}

\author{F. KISLAT$^*$ for the IceCube Collaboration}

\address{DESY\\
D-15738 Zeuthen, Germany\\
$^*$E-mail: fabian.kislat@desy.de}

\begin{abstract}
The IceTop air shower array is the surface component of the IceCube Neutrino Observatory at the geographic South Pole. 
The combination of IceTop and IceCube provides a new and powerful tool to measure cosmic ray composition in the energy range between about~$300\un{TeV}$ and~$1\un{EeV}$ by detecting the electromagnetic component at the surface in coincidence with the muon bundle in the deep underground detector. 
The paper will give an overview of the current status of the detector and the first physics results will be presented.
\end{abstract}


\bodymatter

\section{Introduction}
IceTop is an air shower array and the surface component of the IceCube Neutrino Observatory located at the geographic South Pole\cite{achterberg06}.
It comprises~81 detector stations covering an area of~$1\un{km^2}$ above the neutrino telescope in the ice.
Construction of IceCube and IceTop was completed in the~2010/11 austral summer\cite{kolanoski11}.

The main purpose of IceTop is the measurement of the cosmic ray energy spectrum and chemical composition in the energy range from~$10^{14}$ to~$10^{18}\un{eV}$.
At the so-called ``knee'' at an energy of about~$4\cdot 10^{15}\un{eV}$ the energy spectrum steepens from a spectral index of about~$-2.7$ to about~$-3.1$.
Many models\cite{hoerandel04} predict that this steepening is accompanied by a change of the chemical composition of cosmic rays in the energy range above the knee.
A good measurement of the composition and spectrum in this energy range is thus crucial in order to understand the acceleration mechanisms and the propagation of cosmic rays.
However, the mass determination is notoriously difficult because measurements are indirect and thus afflicted with large systematic errors.
While several experiments have already observed a change in composition above the knee, details of the features remain unclear, reducing discriminative power.

The combination of IceTop and IceCube offers the unique capability to separate the core of high-energy muons from the electromagnetic component of an air shower.
This provides a very powerful way of measuring cosmic ray composition in the PeV energy range.

\section{IceTop detector}
The IceTop stations are located on the surface above IceCube strings, which are instrumented with light detectors at a depth between~$1450\un{m}$ and~$2450\un{m}$. 
They are arranged on a triangular grid with a nominal spacing of~$125\un{m}$.
Three stations are located at intermediate positions at the center of the array forming an infill with a smaller spacing in order to reduce the detector threshold.
Each station consists of two tanks with a diameter of~$1.82\un{m}$ filled with ice to a height of~$90\un{cm}$.
In each tank, Cherenkov light emitted by relativistic charged particles traversing the ice is recorded by two ``Digital Optical Modules'' (DOMs)\cite{DOMPaper}.

The DOMs consist of a photomultiplier tube and electronic circuitry for readout and digitization.
The two DOMs inside an IceTop tank are operated at two different gains (high gain:~$5 \cdot 10^6$; low gain:~$10^5$) in order to increase the linear dynamic range of the tank.
After a trigger, a DOM records and digitizes the PMT signal for about~$422\un{ns}$ with a sampling rate of~$300\un{MSPS}$.
The two tanks of a station are operated in a local coincidence mode requiring both high-gain DOMs to trigger in order to reduce the trigger rate.
Additionally, for all DOM triggers a simple charge and time stamp are recorded (soft local coincidence, SLC).

Near-vertical muons with GeV energies leave a distinct peak in the charge distribution recorded by a tank when no local coincidence is required.
The position of this peak is referred to as~1~Vertical Equivalent Muon~(VEM) and is used to calibrate the tank signals.
All signal charges are thus expressed in units of~VEM.

\section{Air shower reconstruction}
The main primary energy sensitive observable in IceTop is the shower size~$S_{125}$.
It is determined by fitting the lateral distribution of charges with a custom lateral distribution function\cite{StefanThesis}.
The position of the shower core is determined from the lateral charge distribution in the same fit, whereas the shower direction is reconstructed in a fit of the signal times to a function describing the shower front.
With this reconstruction a core position resolution of about~$8\un{m}$ and an angular resolution better than~$0.5^\circ$ was obtained for sufficiently high energies with the 26-station configuration of the detector\cite{FabianPhdThesis}.

\section{Spectrum and composition of cosmic rays}
Like all indirect measurements, IceTop has to rely on Monte Carlo simulations in order to relate the measured shower parameters to the properties of the primary particle.
Therefore, it is important to exploit several systematically independent composition sensitive observables.

\subsection{Coincident events in IceTop and IceCube}
Due to the interplay of decay and interaction of charged pions during the first few interaction lengths, the multiplicity of high-energy muons is larger for showers initiated by heavier primaries.
Thus, measuring the number of TeV muons in IceCube in coincidence with the shower size in IceTop is a very powerful way to measure primary mass.

\begin{figure}[t]
  \psfig{file=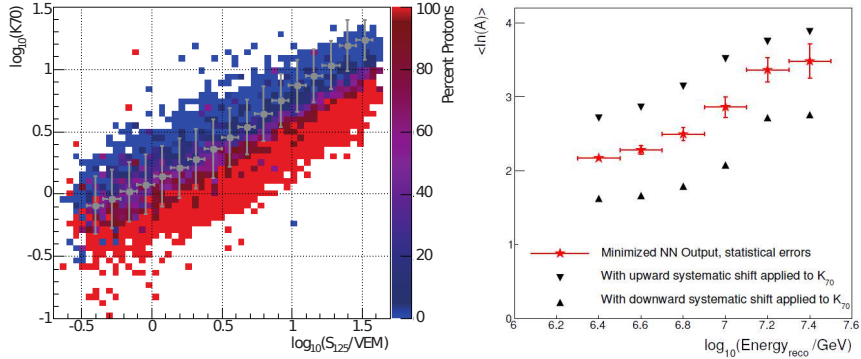,width=\textwidth}
  \caption{Left: Muon bundle size as a function of the shower size at the surface for protons and iron. The numbered lines indicated lines of constant primary energy. Right: Mean logarithmic mass determined from IceCube-40 data.}
  \label{fig:em-vs-mu}
\end{figure}

Figure~\ref{fig:em-vs-mu} (left) shows a simulation of the correlation between the parameter $K_{70}$, which measures the muon energy at a distance of~$70\un{m}$ from the bundle axis in the deep ice, and the electromagnetic component shower size $S_{125}$.
Different primary masses populate different bands in this plot, and the shading indicates the proton content in the simulation: dark grey is~$100\%$ iron, light grey is~$100\%$ protons.
The relation between the $K_{70}-S_{125}$ space and the mass-energy space is non-linear.
The measured data are overlayed.

A neural network has been used to derive the cosmic-ray energy spectrum and mass composition from data taken with the 40 stations and 40 strings configuration of IceCube\cite{rawlins11}.
The resulting energy spectrum is in good agreement with previous results from other experiments and the measured dependence of mean logarithmic mass on energy (Fig.~\ref{fig:em-vs-mu}, right) can be described with the polygonato model\cite{hoerandel03}.

\subsection{Muons in IceTop}
The muon energy distribution in an air shower peaks at a few GeV.
These muons are produced at a much later stage of shower development than the high-energy muons that can reach IceCube.
Distinguishing the signals produced by individual particles in IceTop tanks is not possible.
However, because muons in the GeV range always create a signal of about~$1\un{VEM}$ they can be detected at the periphery of an air shower where the signal expectation value from the electromagnetic component is much smaller than~$1\un{VEM}$.
The abundance of these muons depends on the primary mass.
Thus, they can be exploited to determine the mass composition, which is currently being investigated.

\begin{figure}[t]
  \psfig{file=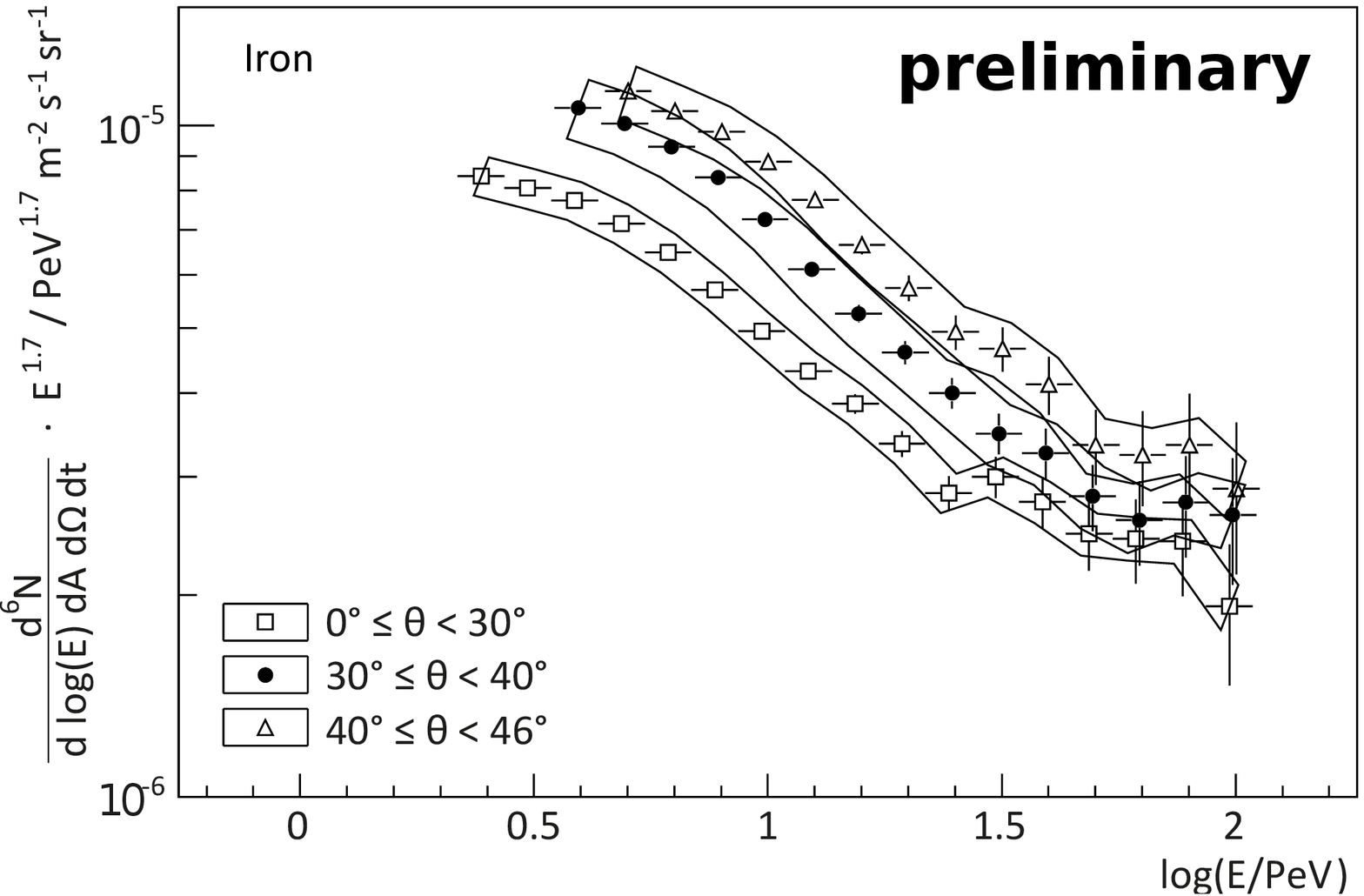,width=.5\textwidth}%
  \psfig{file=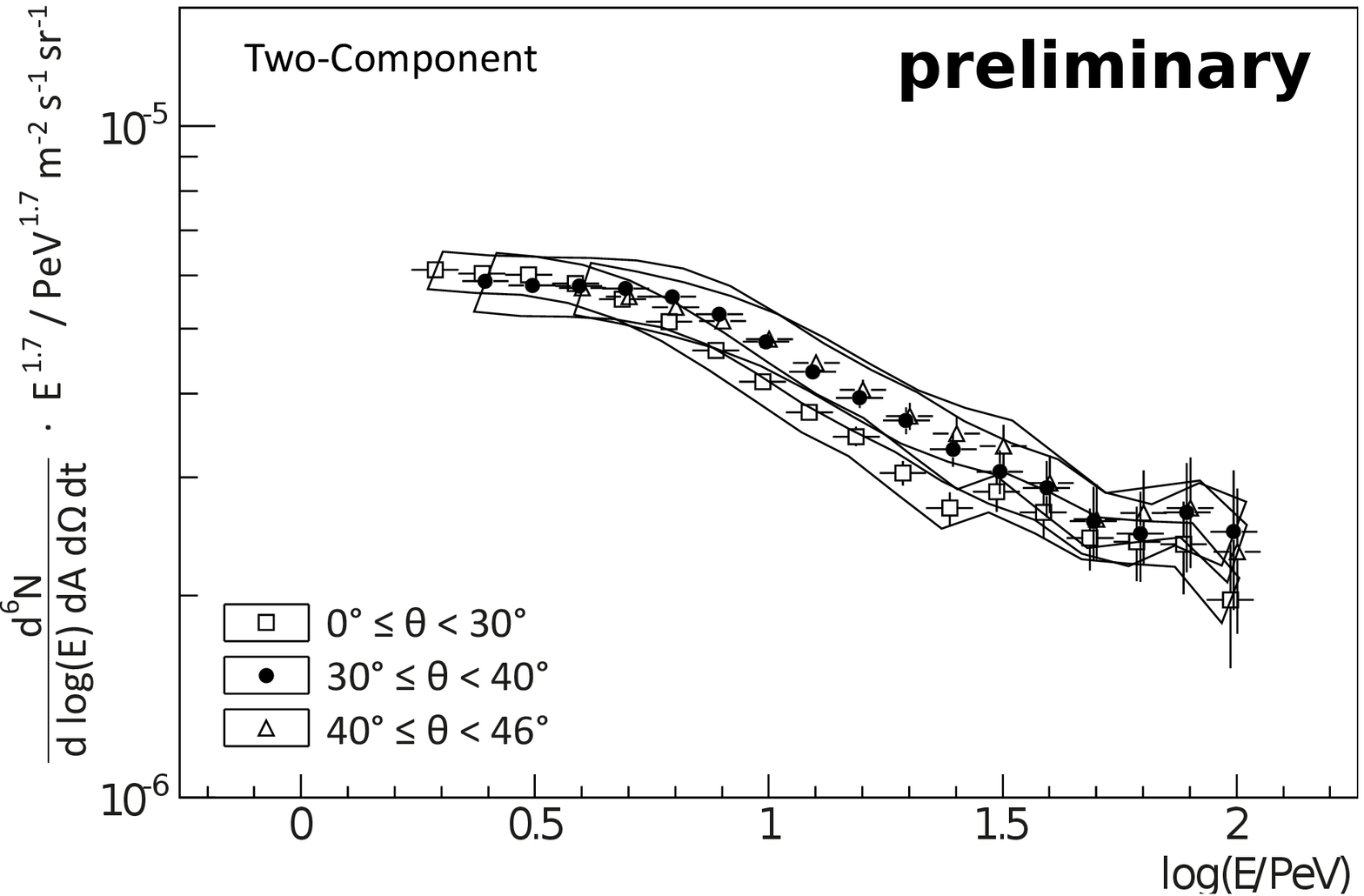,width=.5\textwidth}%
  \caption{Unfolded energy spectra in three different zenith angle ranges, assuming pure iron (left) and a mixture of protons and iron (right). Under the assumption of an isotropic cosmic ray flux, experimental data are in disagreement with pure iron below~$15\un{PeV}$ at a~$99\%$ confidence level.}
  \label{fig:spectra}
\end{figure}

\subsection{Inclined showers}
Showers initiated by heavy primaries develop faster in the atmosphere than those initiated by lighter particles.
Studying showers at various zenith angles allows one to sample the average longitudinal development of air showers at several slant depths with the same detector.
This has been exploited by determining energy spectra from three different zenith angle ranges (see Fig.~\ref{fig:spectra}).
Since cosmic rays can be assumed to be isotropic to a very high degree, the spectra in different zenith angles have to agree.
In the analysis, this has been used to determine the all-particle cosmic-ray energy spectrum and limit the range of potential assumptions on the primary composition\cite{FabianPhdThesis}.

\section{Other results and conclusions}
Since May 2011 IceTop and IceCube are operated in their final configuration with~86 strings and~81 stations.
First results on composition and energy spectrum of cosmic rays are available, while further composition-sensitive observables are being investigated.
In future, combining several measurements of the spectrum and composition of cosmic rays will allow us to reduce systematic errors and to make conclusions about the physics of cosmic-ray interactions in the atmosphere.
Besides its high-energy cosmic-ray physics program IceTop also allows the study of heliospheric events through the variation of individual DOM trigger rates\cite{abbasi08}.
Furthermore, a search for air showers without muons in IceCube allowed setting a limit on the relative abundance of PeV photons\cite{buitink11}.
A study of the anisotropy of cosmic rays is currently in progress.

\bibliographystyle{ws-procs9x6}
\bibliography{bibliography}

\end{document}